\documentclass[a4paper, amsfonts, amssymb, amsmath, reprint, prl, aps,
         showkeys, nofootinbcib, superscriptaddress, twoside, twocolumn]{revtex4-2}

\usepackage{lipsum}
\usepackage{amsthm}
\usepackage{mathtools}
\usepackage{physics}
\usepackage{xcolor}
\usepackage{graphicx}
\usepackage[left=23mm,right=13mm,top=35mm,columnsep=15pt]{geometry} 
\usepackage{adjustbox}
\usepackage{placeins}
\usepackage[T1]{fontenc}
\usepackage{lipsum}
\usepackage[english]{babel}
\usepackage[utf8]{inputenc}
\usepackage{csquotes}
\usepackage[mathscr]{eucal} % used for scripted S symbol
\usepackage{bibunits} % for multiple bibliographies
\usepackage[pdftex, pdftitle={Article}, pdfauthor={Author}]{hyperref} % For hyperlinks in the PDF
\usepackage[capitalize]{cleveref}
\usepackage{multirow}

\newcommand{\taur}{t_{\mathrm{r}}}
\newcommand{\taup}{t_{\mathrm{p}}}
\newcommand{\Done}{\mathrm{D}_{1}}
\newcommand{\Dtwo}{\mathrm{D}_{2}}
\newcommand{\Ancilla}{\mathrm{A}}
\newcommand{\Tone}{T_\mathrm{1}}
\newcommand{\Ttwoecho}{T_\mathrm{2}}
\newcommand{\kHz}{\mathrm{kHz}}
\newcommand{\MHz}{\mathrm{MHz}}
\newcommand{\GHz}{\mathrm{GHz}}
\newcommand{\ns}{\mathrm{ns}} 
\newcommand{\us}{\mathrm{\mu s}}

\begin{document}
% \linenumbers
% Title
\title{All-microwave leakage reduction units for quantum error correction with superconducting transmon qubits}
% Authors
\author{J.~F.~Marques}
\thanks{These two authors contributed equally}
\affiliation{QuTech, Delft University of Technology, P.O. Box 5046, 2600 GA Delft, The Netherlands}
\affiliation{Kavli Institute of Nanoscience, Delft University of Technology, P.O. Box 5046, 2600 GA Delft, The Netherlands}

\author{H.~Ali}
\thanks{These two authors contributed equally}
\affiliation{QuTech, Delft University of Technology, P.O. Box 5046, 2600 GA Delft, The Netherlands}
\affiliation{Kavli Institute of Nanoscience, Delft University of Technology, P.O. Box 5046, 2600 GA Delft, The Netherlands}

\author{B.~M.~Varbanov}
\affiliation{QuTech, Delft University of Technology, P.O. Box 5046, 2600 GA Delft, The Netherlands}

\author{M.~Finkel}
\affiliation{QuTech, Delft University of Technology, P.O. Box 5046, 2600 GA Delft, The Netherlands}
\affiliation{Kavli Institute of Nanoscience, Delft University of Technology, P.O. Box 5046, 2600 GA Delft, The Netherlands}

\author{H.~M.~Veen}
\affiliation{QuTech, Delft University of Technology, P.O. Box 5046, 2600 GA Delft, The Netherlands}
\affiliation{Kavli Institute of Nanoscience, Delft University of Technology, P.O. Box 5046, 2600 GA Delft, The Netherlands}

\author{S.~L.~M.~van~der~Meer}
\affiliation{QuTech, Delft University of Technology, P.O. Box 5046, 2600 GA Delft, The Netherlands}
\affiliation{Kavli Institute of Nanoscience, Delft University of Technology, P.O. Box 5046, 2600 GA Delft, The Netherlands}

\author{S.~Valles-Sanclemente}
\affiliation{QuTech, Delft University of Technology, P.O. Box 5046, 2600 GA Delft, The Netherlands}
\affiliation{Kavli Institute of Nanoscience, Delft University of Technology, P.O. Box 5046, 2600 GA Delft, The Netherlands}

\author{N.~Muthusubramanian}
\affiliation{QuTech, Delft University of Technology, P.O. Box 5046, 2600 GA Delft, The Netherlands}
\affiliation{Kavli Institute of Nanoscience, Delft University of Technology, P.O. Box 5046, 2600 GA Delft, The Netherlands}

\author{M.~Beekman}
\affiliation{QuTech, Delft University of Technology, P.O. Box 5046, 2600 GA Delft, The Netherlands}
\affiliation{Netherlands Organisation for Applied Scientific Research (TNO), P.O. Box 96864, 2509 JG The Hague, The Netherlands}

\author{N.~Haider}
\affiliation{QuTech, Delft University of Technology, P.O. Box 5046, 2600 GA Delft, The Netherlands}
\affiliation{Netherlands Organisation for Applied Scientific Research (TNO), P.O. Box 96864, 2509 JG The Hague, The Netherlands}

\author{B.~M.~Terhal}
\affiliation{QuTech, Delft University of Technology, P.O. Box 5046, 2600 GA Delft, The Netherlands}
\affiliation{EEMCS Department, Delft University of Technology, P.O. Box 5046, 2600 GA Delft, The Netherlands}

\author{L.~DiCarlo}
\email{Corresponding author: l.dicarlo@tudelft.nl}
\affiliation{QuTech, Delft University of Technology, P.O. Box 5046, 2600 GA Delft, The Netherlands}
\affiliation{Kavli Institute of Nanoscience, Delft University of Technology, P.O. Box 5046, 2600 GA Delft, The Netherlands}

% Date
\date{\today}

\begin{bibunit}[naturemag]
% Abstract
\begin{abstract}
Minimizing leakage from computational states is a challenge when using many-level systems like superconducting quantum circuits as qubits. We realize and extend the quantum-hardware-efficient, all-microwave leakage reduction unit (LRU) for transmons in a circuit QED architecture proposed by Battistel et al. This LRU effectively reduces leakage in the second- and third-excited transmon states with up to $99\%$ efficacy in $220~\ns$, with minimum impact on the qubit subspace. As a first application in the context of quantum error correction, we show how multiple simultaneous LRUs can reduce the error detection rate and suppress leakage buildup within $1\%$ in data and ancilla qubits over 50 cycles of a weight-2 stabilizer measurement.
\end{abstract}
\maketitle
% Introduction
\section{Introduction} \label{sec:introduction}

Superconducting qubits, such as the transmon~\cite{koch07}, are many-level systems in which a qubit is represented by the two lowest-energy states $\ket{g}$ and $\ket{e}$. However, leakage to non-computational states is a risk for all quantum operations, including single-qubit gates~\cite{Motzoi09}, two-qubit gates~\cite{DiCarlo09, Barends19, Negirneac21} and measurement~\cite{Sank16,Khezri22}. While the typical probability of leakage per operation may pale in comparison to conventional qubit errors induced by control errors and decoherence~\cite{Negirneac21, Chen21}, unmitigated leakage can build up with increasing circuit depth. A prominent example is multi-round quantum error correction (QEC) with stabilizer codes such as the surface code~\cite{Fowler12}. In the absence of leakage, such codes successfully discretize all qubit errors into Pauli errors through the measurement of stabilizer operators~\cite{Riste15,Takita17}, and these Pauli errors can be detected and corrected (or kept track of) using a decoder. However, leakage errors fall outside the qubit subspace and are not immediately correctable~\cite{Aliferis07, Fowler13, Ghosh13_b}. The signature of leakage on the stabilizer syndrome is often not straightforward, hampering the ability to detect and correct it~\cite{Varbanov20,Bultink20}. Additionally, the build-up of leakage over QEC rounds accelerates the destruction of the logical information~\cite{Mcewen21, Chen21}.
Therefore, despite having low probability per operation, methods to reduce leakage must be employed when performing experimental QEC with multi-level systems.

Physical implementations of QEC codes~\cite{Ryan-Anderson21, krinner22, Zhao22, Sundaresan22, Acharya22, Postler22} use qubits for two distinct functions: Data qubits store the logical information and, together, comprise the encoded logical qubits. Ancilla qubits perform indirect measurement of the stabilizer operators. Handling leakage in ancilla qubits is relatively straightforward as they are measured in every QEC cycle. This allows for the use of reset protocols~\cite{Magnard18,Mcewen21} without the loss of logical information. Leakage events can also be directly detected using three- or higher-level readout~\cite{krinner22} and reset using feedback~\cite{Riste12, Andersen19}. In contrast, handling data-qubit leakage requires a subtle approach as it cannot be reset nor directly measured without loss of information or added circuit complexity~\cite{Ghosh15, Suchara15, Mcewen23}. A promising solution is to interleave QEC cycles with operations that induce seepage without disturbing the qubit subspace, known as leakage reduction units (LRUs)~\cite{Aliferis07, Fowler13, Ghosh15, Suchara15, Brown19, Brown20, Hayes20, Battistel21, Miao22}.
An ideal LRU returns leakage back to the qubit subspace, converting it into Pauli errors which can be detected and corrected, while leaving qubit states undisturbed. By converting leakage into conventional errors, LRUs enable a moderately high physical noise threshold, below which the logical error rate decreases exponentially with the code distance~\cite{Fowler13, Suchara15}.
A more powerful operation called 'heralded leakage reduction' would both reduce and herald leakage, leading to a so-called erasure error~\cite{Grassl97,Bennett97}.
Unlike Pauli errors, the exact location of erasures is known, making them easier to correct and leading to higher error thresholds~\cite{Stace09, Barrett10, Kubica22, Wu2022}.

In this Letter, we present the realization and extension of the LRU scheme proposed in Ref.~\onlinecite{Battistel21}. This is a highly practical scheme requiring only microwave pulses and the quantum hardware typically found in contemporary circuit QED quantum processors: a microwave drive and a readout resonator dispersively coupled to the target transmon (in our case, a readout resonator with dedicated Purcell filter). We show its straightforward calibration and the effective removal of the population in the first two leakage states of the transmon ($\ket{f}$ and $\ket{h}$) with up to $>99\%$ efficacy in $220~\ns$. Process tomography reveals that the LRU backaction on the qubit subspace is only an AC-Stark shift, which can be easily corrected using a $Z$-axis rotation. As a first application in a QEC setting, we interleave repeated measurements of a weight-2 parity check~\cite{Andersen19, Bultink20} with simultaneous LRUs on data and ancilla qubits, showing the suppression of leakage and error detection rate buildup.

% Results
\section{Results}
\label{sec:results}

\begin{figure}
\centering
\includegraphics[width=0.49\textwidth]{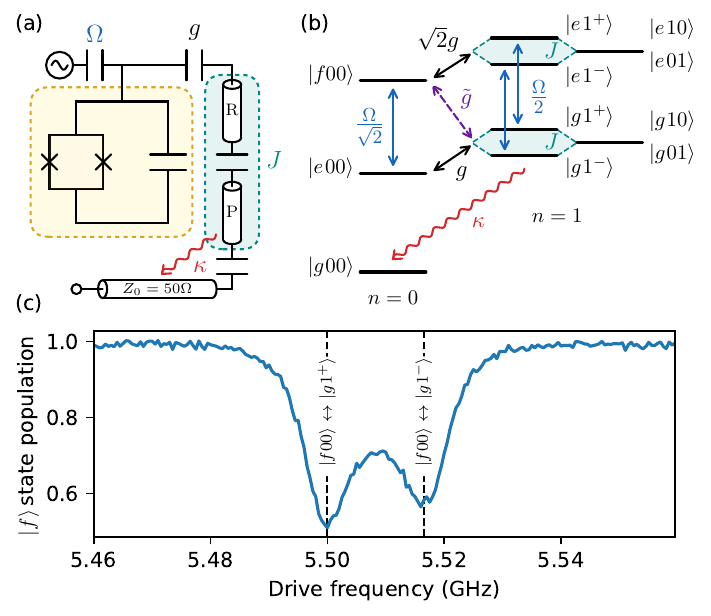}
\caption{\textbf{Leakage reduction unit scheme.}
(a) Schematic for the driven transmon-resonator system. A transmon ($\mathrm{T}$, yellow) with three lowest-energy levels $\ket{g}$, $\ket{e}$, and $\ket{f}$  is coupled to a readout resonator $(\mathrm{R})$ with strength $g$. The latter is coupled to a frequency-matched Purcell resonator $(\mathrm{P})$ with strength $J$.  The Purcell resonator also couples to a $50~\Omega$ feedline through which its excitations quickly decay at rate $\kappa$.
The transmon is driven with a pulse of strength $\Omega$ applied to its microwave drive line.
(b) Energy level-spectrum of the system. Levels are denoted as $\ket{\mathrm{T},\mathrm{R},\mathrm{P}}$, with numbers indicating photons in $\mathrm{R}$ and $\mathrm{P}$.
As the two resonators are frequency matched, the right-most degenerate states split by $2J$, and $g$ is shared equally among the two hybridized resonator modes $\ket{1^-}$ and $\ket{1^+}$.
An effective coupling $\tilde{g}$ arises between $\ket{f00}$ and the two hybridized states $\ket{g1^\pm}$ via $\ket{e00}$ and $\ket{e1^\pm}$.
(c) Spectroscopy of the $\ket{f00}\leftrightarrow\ket{g1^\pm}$ transition. Measured transmon population in $\ket{f}$ versus drive frequency, showing dips corresponding to the two transitions assisted by each of the hybridized resonator modes.}
\label{fig:lru_scheme}
\end{figure}

Our leakage reduction scheme [Fig.~\ref{fig:lru_scheme}(a)] consists of a transmon with states $\ket{g}$, $\ket{e}$ and $\ket{f}$, driven by an external drive $\Omega$, coupled to a resonant pair of Purcell and readout resonators~\cite{Heinsoo18} with effective dressed states $\ket{00}$ and $\ket{1^\pm}$.
The LRU scheme transfers leakage population in the second-excited state of the transmon, $\ket{f}$, to the ground state, $\ket{g}$, via the resonators using a microwave drive.
It does so using an effective coupling $\tilde{g}$ mediated by the transmon-resonator coupling, $g$, and the drive $\Omega$, which couples states $\ket{f00}$ and $\ket{g1^\pm}$.
Driving at the frequency of this transition,
\begin{equation}
\omega_{f00}-\omega_{g1^\pm} \approx 2\omega_{\mathrm{Q}}+\alpha-\omega_{\mathrm{RP}},
\label{eq:tansition-freq}
\end{equation}
transfers population from $\ket{f00}$ to $\ket{g1^\pm}$, which in turn quickly decays to $\ket{g00}$ provided the Rabi rate is small compared to $\kappa$. Here, $\omega_{\mathrm{Q}}$ and $\alpha$ are the transmon qubit transition frequency and anharmonicity, respectively, while $\omega_{\mathrm{RP}}$ is the resonator mode frequency.
In this regime, the drive effectively pumps any leakage in $\ket{f}$ to the computational state $\ket{g}$.
We perform spectroscopy of this transition by initializing the transmon in $\ket{f}$ and sweeping the drive around the expected frequency.
The results [Fig.~\ref{fig:lru_scheme}(c)] show two dips in the $f$-state population corresponding to transitions with the hybrized modes of the matched readout-Purcell resonator pair.
The dips are broadened by $\sim\kappa/4\pi\approx8~\MHz$, making them easy to find.
\begin{figure}
\centering
\includegraphics[width=0.49\textwidth]{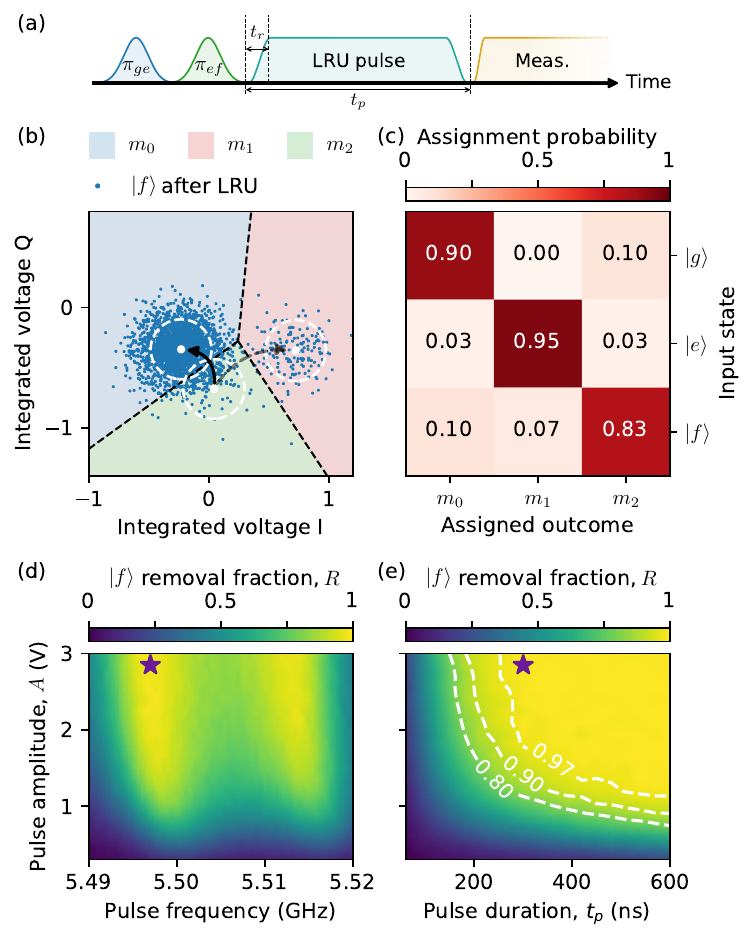}
\caption{\textbf{Calibration of the leakage reduction unit pulse.}
(a) Pulse sequence used for LRU calibration.
(b) Single-shot readout data obtained from the experiment. The blue, red and green areas denote $m_0$, $m_1$, and $m_2$ assignment regions, respectively. The mean (white dots) and $3\sigma$ standard deviation (white dashed circles) shown are obtained from Gaussian fits to the three input-state distributions. The blue data shows the first $3\times10^{3}$ (from a total of $2^{15}$) shots of the experiment described in (a), indicating $99.(3)\%$ $\ket{f}$-state removal fraction.
(c) Measured assignment fidelity matrix used for readout correction.
(d-e) Extracted $\ket{f}$-state removal fraction versus pulse parameters. Added contours (white dashed curves) indicate  $80$, $90$ and $97\%$ removal fraction. The purple star indicates the pulse parameters used in (b).}
\label{fig:gate_cal}
\end{figure}

To make use of this scheme for a LRU, we calibrate a pulse that can be used as a circuit-level operation.
We use the pulse envelope proposed in Ref.~\onlinecite{Battistel21}:
\begin{equation}
A(t) =
\begin{cases}
A\sin^2\left(\pi\frac{t}{2\taur}\right) & \text{for $0\leq t\leq \taur$,}\\
A & \text{for $\taur\leq t\leq \taup-\taur$,}\\
A\sin^2\left(\pi\frac{\taup-t}{2\taur}\right) & \text{for $\taup-\taur\leq t\leq \taup$,}
\end{cases}
\label{eq:pulse_param}
\end{equation}
where $A$ is the amplitude, $\taur$ is the rise and fall time, and $\taup$ is the total duration.
We conservatively choose $\taur=30~\ns$ to avoid unwanted transitions in the transmon.
To measure the fraction of leakage removed, $R$, we apply the pulse on the transmon prepared in $\ket{f}$ and measure it [Fig.~\ref{fig:gate_cal}(a)], correcting for readout error using the measured 3-level assignment fidelity matrix [Fig.~\ref{fig:gate_cal}(c)].
To optimize the pulse parameters, we first measure $R$ while sweeping the pulse frequency and $A$ [Fig.~\ref{fig:gate_cal}(d)].
A second sweep of $\taup$ and $A$ [Fig.~\ref{fig:gate_cal}(e)] shows that $R>99\%$ can be achieved by increasing either parameters.
Simulation~\cite{Battistel21} suggests that $R\approx80\%$ is already sufficient to suppress most of the impact of current leakage rates, which is comfortably achieved over a large region of parameter space.  For QEC, a fast operation is desirable to minimize the impact of decoherence.
However, one must not excessively drive the transmon, which can cause extra decoherence (see Fig.~6 in Ref.~\onlinecite{Battistel21}).
Considering the factors above, we opt for $\taup=220~\ns$ and adjust $A$ such that $R\gtrsim80\%$.
Additionally, we benchmark the repeated action of the LRU and verify that its performance is maintained over repeated applications, thus restricting leakage events to approximately a single cycle (Fig.~\ref{fig:repeated_LRU_experiment}).

\begin{figure}
\centering
\includegraphics[width=0.49\textwidth]{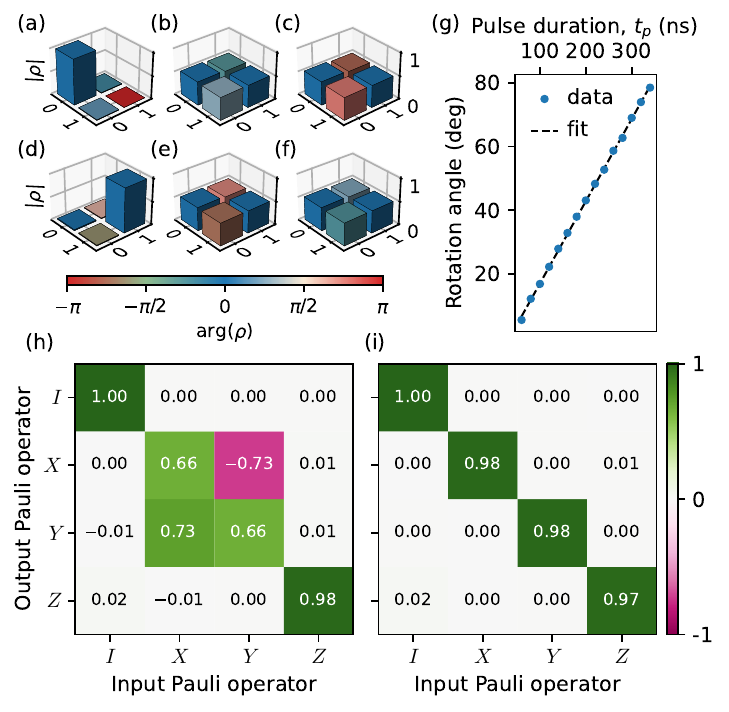}
\caption{\textbf{Process tomography of the leakage reduction unit.}
(a-f) Measured density matrices after the LRU gate for input states $\ket{0}$, $\ket{+}$, $\ket{+i}$, $\ket{1}$, $\ket{-}$, and $\ket{-i}$, respectively.
(g) $Z$-rotation angle induced on the qubit versus the LRU pulse duration. The linear best fit (black dashed line) indicates an AC-Stark shift of $71(9)$ kHz.
(h-i) Pauli transfer matrix of the LRU with (i) and without (h) virtual phase correction ($\taup=220~\ns$ and $R=84.(7)\%$).}
\label{fig:proc_tomo}
\end{figure}

With the LRU calibrated, we next benchmark its impact on the qubit subspace using quantum process tomography.
The results (Fig.~\ref{fig:proc_tomo}) show that the qubit incurs a $Z$-axis rotation.
We find that the rotation angle increases linearly with $\taup$ [Fig.~\ref{fig:proc_tomo}(g)], consistent with a  $71(9)~\kHz$ AC-Stark shift induced by the LRU drive.
This phase error in the qubit subspace can be avoided using decoupling pulses or corrected with a virtual $Z$ gate.
Figures~\ref{fig:proc_tomo}(h) and~\ref{fig:proc_tomo}(i) show the Pauli transfer matrix (PTM) for the operation before and after applying a virtual $Z$ correction, respectively.
From the measured PTM [Fig.~\ref{fig:proc_tomo}(i)] and enforcing physicality constraints~\cite{Chow12}, we obtain an average gate fidelity $F_\mathrm{avg}=98.(9)\%$.
Compared to the measured  $99.(2)\%$ fidelity of idling during the same time ($\taup=220~\ns$), there is evidently no significant error increase.

\begin{figure}
\centering
\includegraphics[width=0.49\textwidth]{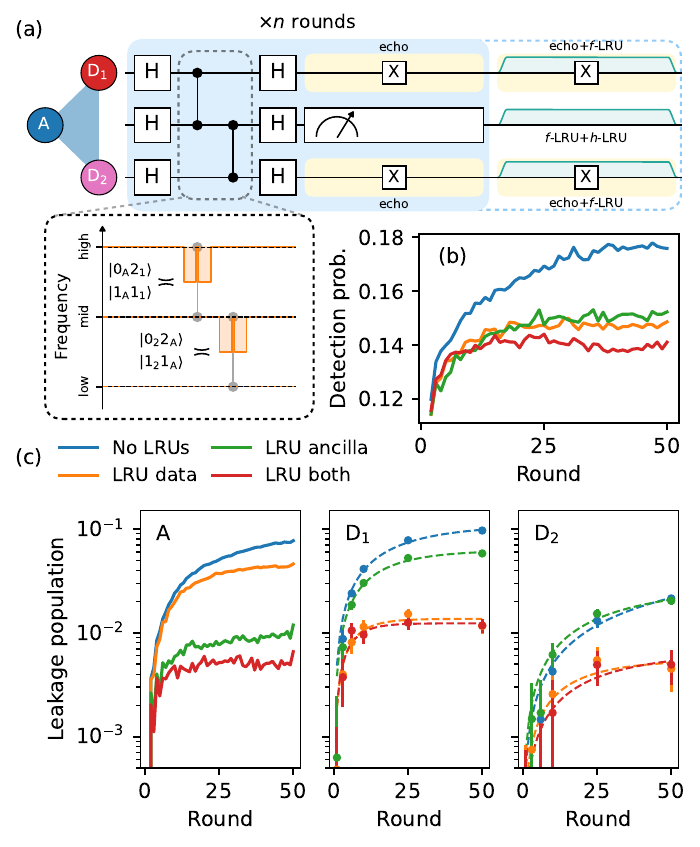}
\caption{\textbf{Repeated stabilizer measurement with leakage reduction.}
(a) Quantum circuit using ancilla $\Ancilla$ to measure the $X$-type parity of data qubits $\Done$ and $\Dtwo$.
The dashed box shows the frequency arrangement for two-qubit CZ gates.
A CZ gate is performed by fluxing the higher-frequency transmon down in frequency to the nearest avoided crossing (orange shaded trajectory).
The duration of single-qubit gates, CZ gates, and measurement are $20$, $60$ and $340~\ns$, respectively, totalling $500~\ns$ for the parity check (light-blue region).
Performing the LRUs extends the circuit by $220~\ns$ (blue-dashed region).
Echo pulses on data qubits mitigate phase errors caused by residual $ZZ$ crosstalk and AC-stark shift during the measurement and LRUs (light yellow slots).
(b-c) Measured error-detection probability (b) and leakage (c) versus the number of parity-check rounds in four settings.
The No LRUs setting (blue) does not apply any LRUs.
LRU data (orange) and LRU ancilla (green) settings apply LRUs exclusively on the data qubits and the ancilla, respectively.
The LRU both (red) setting applies LRUs on all qubits.
}
\label{fig:repeated_parity_check}
\end{figure}

Finally, we implement the LRU in a QEC scenario by performing repeated stabilizer measurements of a weight-2 $X$-type parity check~\cite{Andersen19, Bultink20} using three transmons (Fig.~\ref{fig:repeated_parity_check}).
We use the transmon in Figs.~1-3, $\Done$, plus an additional transmon ($\Dtwo$) as data-qubits together with an ancilla, $\Ancilla$.
LRUs for $\Dtwo$ and $\Ancilla$ are tuned using the same procedure as above.
A detailed study of the performance of this parity check is shown in the Supplementary Information (Fig.~\ref{fig:parity-bench}).
Given their frequency configuration~\cite{Versluis17}, $\Done$ and $\Ancilla$ are most vulnerable to leakage during two-qubit controlled-$Z$ (CZ) gates, as shown by the avoided crossings in Fig.~\ref{fig:repeated_parity_check}(a). Additional leakage can occur during other operations: in particular, we observe that leakage into states above $\ket{f}$ can occur in $\Ancilla$ due to measurement-induced transitions~\cite{Sank16} (see Fig.~\ref{fig:anc_leakage}). Therefore, a LRU acting on $\ket{f}$ alone is insufficient for $\Ancilla$.
To address this, we develop an additional LRU for $\ket{h}$ ($h$-LRU), the third-excited state of $\Ancilla$ (see Supplementary Information Fig.~\ref{fig:h-LRU}).
The $h$-LRU can be employed simultaneously with the $f$-LRU without additional cost in time or impact on the $\ket{f}$ removal fraction, $R$.
Thus, we simultaneously employ $f$-LRUs for all three qubits and an $h$-LRU for $\Ancilla$ [Fig.~\ref{fig:repeated_parity_check}(a)].
To evaluate the impact of the LRUs, we measure the error detection probability (probability of a flip occurring in the measured stabilizer parity) and leakage population of the three transmons over multiple rounds of stabilizer measurement.
Without leakage reduction, the error detection probability rises $\sim8\%$ in 50 rounds.
We attribute this feature to leakage build-up~\cite{Mcewen21, Acharya22, Miao22}.
With the LRUs, the rise stabilizes faster (in $\sim10$ rounds) to a lower value and is limited to $2\%$, despite the longer cycle duration ($500$ versus $720~\ns$ without and with the LRU, respectively).
Leakage is overall higher without LRUs, in particular for $\Done$ and $\Ancilla$ [Fig.~\ref{fig:repeated_parity_check}(c)], which show a steady-state population of $\approx10\%$. Using leakage reduction, we lower the leakage steady-state population by up to one order of magnitude to $\lesssim1\%$ for all transmons.
Additionally, we find that removing leakage on other transmons leads to lower overall leakage, suggesting that leakage is transferred between transmons~\cite{Varbanov20, Miao22}.
This is particularly noticeable in $\Ancilla$ [Fig.~\ref{fig:repeated_parity_check}(c)], where the steady-state leakage is always reduced by adding LRUs on $\Done$ and $\Dtwo$.

% Discussion
\section{Discussion} \label{sec:discussion}

We have demonstrated and extended the all-microwave LRU for superconducting qubits in circuit QED proposed in Ref.~\onlinecite{Battistel21}.
We have shown how these LRUs can be calibrated using a straightforward procedure to deplete leakage in the second- and third-excited states of the transmon.
This scheme could potentially work for even higher transmon states using additional drives.
We have verified that the LRU operation has minimal impact in the qubit subspace, provided one can correct for the static AC-Stark shift induced by the drive(s).

This scheme does not reset the qubit state and is therefore compatible with both data and ancilla qubits in the QEC context.
We have showcased the benefit of the LRU in a building-block QEC experiment where LRUs decrease the steady-state leakage population of data and ancilla qubits by up to one order of magnitude (to $\lesssim1\%$), and thereby reduce the error detection probability of the stabilizer and reaching a faster steady state. 
We find that the remaining ancilla leakage is dominated by higher states above $\ket{f}$ (Fig.~\ref{fig:anc_leakage}) likely caused by the readout~\cite{Sank16, Khezri22}.
Compared to other LRU approaches ~\cite{Mcewen21, Miao22},  we believe this scheme is especially practical as it is all-microwave and very quantum-hardware efficient, requiring only the microwave drive line and dispersively coupled resonator that are already commonly found in the majority of circuit QED quantum processors~\cite{krinner22, Zhao22, Acharya22}. 
Extending this leakage reduction method to larger QEC experiments can be done without further penalty in time as all LRUs can be simultaneously applied. However, we note that when extending the LRU to many qubits, microwave crosstalk should be taken into account in order to avoid driving unwanted transitions. This can be easily avoided by choosing an appropriate resonator-qubit detuning. 
% Acknowledgements, Competing interests, Data availability
\section{Acknowledgements}
    We thank F.~Battistel and Y. Herasymenko for helpful discussions, and G.~Calusine and W.~Oliver for providing the traveling-wave parametric amplifiers used in the readout amplification chain. This research is supported by the Office of the Director of National Intelligence (ODNI), Intelligence Advanced Research Projects Activity (IARPA), via the U.S. Army Research Office Grant No. W911NF-16-1-0071, by Intel Corporation, and by QuTech NWO funding 2021- 2026 – Part I “Fundamental Research”, project number 601.QT.001-1, financed by the Dutch Research Council (NWO). The views and conclusions contained herein are those of the authors and should not be interpreted as necessarily representing the official policies or endorsements, either expressed or implied, of the ODNI, IARPA, or the U.S. Government.

\section{Competing Interests}
	The authors declare no competing interests.

\section{Data Availability}
	The data supporting the plots and claims within this paper are available online at \url{http://github.com/DiCarloLab-Delft/Leakage_Reduction_Unit_Data}. Further data can be provided upon reasonable request.

% Bibliography
% \putbib[../../Paper_resources/References/References_cQED]
%%%%%%%%%%%%%%%%%%%%%

%%%%%%%%%%%%%%%%%%%%%
\end{bibunit}
% Appendix
\onecolumngrid
\clearpage
\renewcommand{\theequation}{S\arabic{equation}}
\renewcommand{\thefigure}{S\arabic{figure}}
\renewcommand{\thetable}{S\arabic{table}}
\renewcommand{\bibnumfmt}[1]{[S#1]}
\renewcommand{\citenumfont}[1]{S#1}
\setcounter{figure}{0}
\setcounter{equation}{0}
\setcounter{table}{0}
\begin{bibunit}[naturemag]
\section*{Supplemental material for 'All-microwave leakage reduction units for quantum error correction with superconducting transmon qubits'}
% \date{\today}
% \maketitle
\twocolumngrid

This supplement provides additional information in support of the statements and claims in the main text.

\section{Device}
\begin{figure}[b!]
\centering
\includegraphics[width=.49\textwidth]{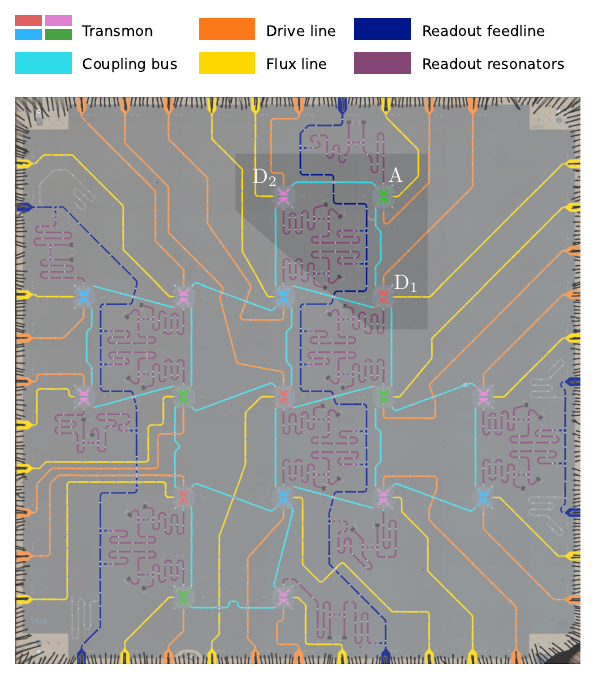}
\caption{\textbf{Circuit QED device.}
Optical image of the 17-transmon quantum processor, with added falsecolor to highlight different circuit elements. The shaded area indicates the three transmons used in this experiment.}
\label{fig:Device}
\end{figure}

\begin{table}[!h]
\begin{tabular}{lccc}
\hline
Transmon                                                 & $\Done$ & $\Ancilla$ & $\Dtwo$  \\
\hline
Frequency at sweetspot, $\omega_\mathrm{Q}/2\pi$ ($\GHz$)  & 6.802          & 6.033        & 4.788           \\
Anharmonicity, $\alpha/2\pi$ ($\MHz$)                      & -295           & -310         & -321            \\
Resonator frequency, $\omega_\mathrm{R}/2\pi$ ($\GHz$)     & 7.786          & 7.600        & 7.105           \\
Purcell res. linewidth, $\kappa/2\pi$ ($\MHz$)             & 15.(5)         & 22.(5)       & 12.(6)          \\
$f$-LRU drive frequency ($\GHz$)                           & 5.498          & 4.135        & 2.152           \\
$h$-LRU drive frequency ($\GHz$)                           & -              & 3.496        & -               \\
$\Tone$ ($\us$)                                         & 17             & 26           & 37              \\
$\Ttwoecho$ ($\us$)                                     & 19             & 22           & 27              \\
Single-qubit gate error (\%)                            & 0.1(0)         & 0.0(7)       & 0.0(5)          \\
Two-qubit gate error (\%)                               & \multicolumn{3}{c}{1.(1)$\:\:\:\:\:\:\:\:$1.(9)}\\
Two-qubit gate leakage (\%)                             & \multicolumn{3}{c}{0.3(7)$\:\:\:\:\:\:$0.1(1)}  \\
$f$-LRU removal fraction, $R^f$ (\%)                    & 84.(7)         & 99.(2)       & 80.(3)          \\
$h$-LRU removal fraction, $R^h$ (\%)                    & -              & 96.(1)       & -               \\
\hline
Operation                                               & \multicolumn{3}{c}{Duration ($\ns$)} \\
\hline
Single-qubit gate                                       & \multicolumn{3}{c}{20}            \\
Two-qubit gate                                          & \multicolumn{3}{c}{60}            \\
Measurement                                             & \multicolumn{3}{c}{340}           \\
LRU                                                     & \multicolumn{3}{c}{220}           \\
\hline
\end{tabular}
\caption{\textbf{Device metrics.} Frequencies and coherence times are measured using standard spectroscopy and time-domain measurements~\cite{Krantz19}. Gate errors are evaluated using randomized benchmarking protocols~\cite{Magesan12, Magesan12b, Wood18}.}
\label{tab:Device}
\end{table}

\begin{figure}[!h]
\centering
\includegraphics[width=.4\textwidth]{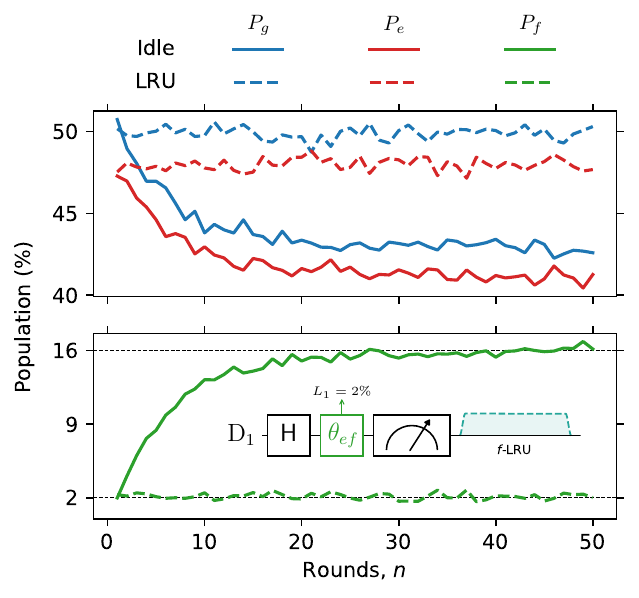}
\caption{\textbf{Repeated LRUs on $\Done$.}
Computational (top) and leakage (bottom) state population over repeated cycles of the circuit as shown.
Transmon $\Done$ is repeatedly put in superposition,  controllably leaked with rate $L_1$ and measured. Measurement is followed by either idling (solid) or the LRU (dashed).
The horizontal lines denote the steady-state leakage with and without LRUs.}
\label{fig:repeated_LRU_experiment}
\end{figure}

\begin{figure*}[!ht]
\centering
\includegraphics[width=.99\textwidth]{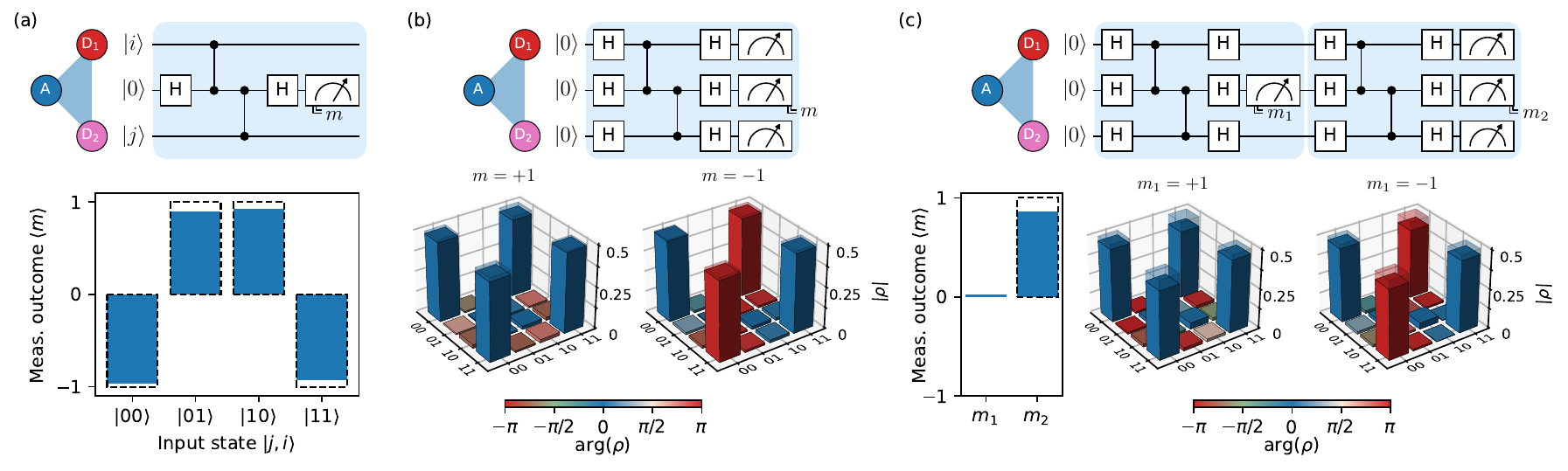}
\caption{\textbf{Benchmarking of the weight-2 parity check.}
(a) Quantum circuit of the weight-2 $X$-type parity check and bar plot of the average measured ancilla outcome for the different input computational states of the data-qubit register.
Dashed bars show ideal average outcome: $\langle m \rangle=+1\, (-1)$ for even (odd) data-qubit input parity.
(b) Generation of Bell states via stabilizer measurement (top) and corresponding data-qubit state tomography (bottom) conditioned on the ancilla outcome. The obtained fidelity to the ideal Bell states (shaded wireframe) is $96.9\%$ and $98.5\%$ for $m=+1$ and $m=-1$, respectively.
(c) Repeated stabilizer experiment. (Bottom left) Average measured ancilla outcome $\langle m\rangle$ for each round of stabilizer measurement. Ideally, the first outcome should be random and the second always $+1$. The measured probability is $P(m_2=+1)=90.0\%$. (Bottom right) Reconstructed data-qubit states conditioned on the first ancilla outcome. The obtained Bell-state fidelities are $90.6\%$ and $91.5\%$ for $m_1=+1$ and $m_1=-1$, respectively.}
\label{fig:parity-bench}
\end{figure*}
The device used (Fig.~\ref{fig:Device}) has 17 flux-tunable transmons arranged in a square lattice with nearest-neighbor connectivity (as required for a distance-3 surface code).
Transmons are arranged in three frequency groups as prescribed in the pipelined architecture of Ref.~\onlinecite{Versluis17}.
Each transmon has a dedicated microwave drive line used for single-qubit gates and leakage reduction, and a flux line used for two-qubit gates.
Nearest-neighbor transmons have fixed coupling mediated by a dispersively coupled bus resonator.
Each transmon has a dedicated pair of frequency-matched readout and Purcell resonators coupled to one of three feedlines, used for fast multiplexed readout in the architecture of Ref.~\onlinecite{Heinsoo18}.
Single-qubit gates are realized using standard DRAG pulses~\cite{Motzoi13}.
Two-qubit controlled-$Z$ gates are implemented using sudden net-zero flux pulses~\cite{Negirneac21}.
Characteristics and performance metrics of the three transmons used in the experiment are shown in Tab.~\ref{tab:Device}.

\section{Repeated LRU application}
For QEC we require that the LRU performance remains constant over repeated applications.
To assess this, we perform repeated rounds of the experiment shown in Fig.~\ref{fig:repeated_LRU_experiment} while idling or using the LRU.
In each round apply an $e$-$f$ rotation with rotation angle $\theta$ to induce a leakage rate
\begin{equation}
L_1 = \frac{\sin^2(\theta/2)}{2},
\end{equation}
and choose $\theta$ such that $L_1=2\%$.
For the purpose of this experiment, we lower the readout amplitude in order to suppress leakage to higher states during measurement~\cite{Sank16},
The results (Fig.~\ref{fig:repeated_LRU_experiment}) show that while idling, leakage in $\ket{f}$ builds up to a steady-state population of about $16\%$.
Using the LRU, it remains constant at $P_f=L_1$ throughout all rounds.
This behavior shows that LRU performance is maintained throughout repeated applications and suggests that leakage events are restricted to a single round.

\section{Benchmarking the weight-2 parity check}
We benchmark the performance of the weight-2 parity check using three experiments assessing different error types.
First, we assess the ability to accurately assign the parity of the data-qubit register by measuring the ancilla outcome for all data-qubit input computational states.
The results [Fig.~\ref{fig:parity-bench}(a)] show an average parity assignment fidelity of 95.6\%.
Next, we look at errors occurring on the data qubits when projecting them onto a Bell state using a $X$-type parity check [Fig.~\ref{fig:parity-bench}(b)].
From data-qubit state tomography conditioned on ancilla outcome, we obtain an average Bell-state fidelity of $97.7\%$ ($96.9\%$ for $m=+1$ and $98.5\%$ for $m=-1$).
For each reported density matrix, we apply readout corrections and enforce physicality constraints via maximum likelihood estimation~\cite{Chow12}.
Finally, we look at the backaction of two back-to-back parity checks [Fig.~\ref{fig:parity-bench}(c)]. Here, we measure the correlation of the two parity outcomes.
Ideally, the first parity outcome should be random while the second should be the same as the first.
Since our ancilla is not reset after measurement, the probability of both parities being correlated is $P(m_2=+1)=90.0\%$ [bar plot in Fig.~\ref{fig:parity-bench}(c)].
We can also reconstruct the data qubit state after the experiment. Here, we find that the average Bell-state fidelity drops to $91.0\%$ ($90.6\%$ for $m_1=+1$ and $91.5\%$ for $m_1=-1$).
This drop in fidelity is likely due to decoherence from idling during the first ancilla measurement.

\section{Measurement-induced transitions}
\begin{figure}
\centering
\includegraphics[width=.48\textwidth]{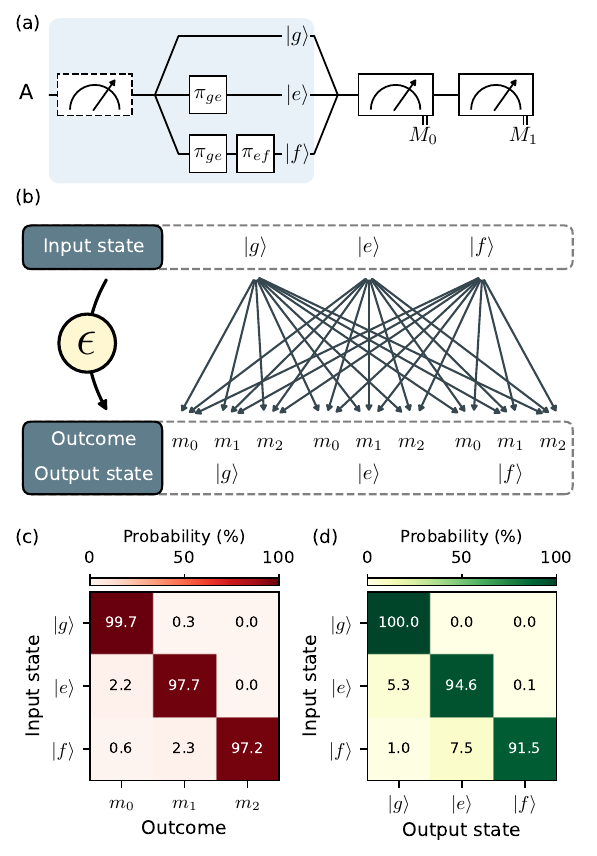}
\caption{\textbf{Characterizing measurement-induced transitions.}
(a) Quantum circuit used to characterize transmon measurement.
A transmon is initialized into states $\ket{g}$, $\ket{e}$ and $\ket{f}$ after a heralding (dashed) measurement (blue panel).
Following prepration, two consecutive measurements $M_0$ and $M_1$ are performed, yielding three-level outcomes.
(b) Illustration of the extracted measurement model. The model is described by a rank $3$ tensor $\epsilon_i^{m, j}$, where input states $i$ are connected to measurement outcomes $m$ and output states $j$.
From it, the assignment probability matrix (c) and the QNDness matrix (d) can be extracted.}
\label{fig:butterfly}
\end{figure}
Previous studies have observed measurement-induced state transitions that can lead to leakage~\cite{Sank16, Khezri22}.
To evaluate the backaction of ancilla measurement, we model the measurement as a rank 3 tensor $\epsilon_i^{m,j}$ which takes an input state $i$, declares an outcome $m$ and outputs a state $j$ with normalization condition,
\begin{equation}
\sum_m \sum_j \epsilon_i^{m, j} = 1.
\end{equation}

To find $\epsilon_i^{m,j}$, we perform the experiment in Fig.~\ref{fig:butterfly}(a).
For each input state $i$, the probability distribution of the measured results $P_i(M_0, M_1)$ follows
\begin{equation}
P_i(M_0=m_k, M_1=m_\ell) = \sum_s\sum_j \epsilon_i^{m_k,s} \epsilon_s^{m_\ell, j}.
\end{equation}
This system of 27 second-order equations is used to estimate the 27 elements of $\epsilon_i^{m, j}$ through a standard optimization procedure.
In this description, the assignment fidelity matrix $M_{i,m}$ [Fig.~\ref{fig:butterfly}(b)] is given by
\begin{equation}
M_{i,m}=\sum_j\epsilon_i^{m,j}.
\end{equation}
Furthermore, this model allows us to assess the probability of transitions occurring during the measurement.
This is given by the QNDness matrix
\begin{equation}
Q_{i,j}=\sum_m\epsilon_i^{m,j}.
\end{equation}
The results [Fig.~\ref{fig:butterfly}(b)] show an average QNDness of $95.4\%$ across all states.
The average leakage rate ($(Q_{g,f}+Q_{e,f})/2$) is $0.06\%$, predominantly occurring for input state $\ket{e}$.

% Mention QNDness ($\ket{g}$ $100.0\%$, $\ket{e}$ $94.6\%$, $\ket{f}$ $91.5\%$ avg $95.4\%$) and leakage rate ($0.1\%$).

\section{Readout of transmon states}

\begin{figure}
\centering
\includegraphics[width=.49\textwidth]{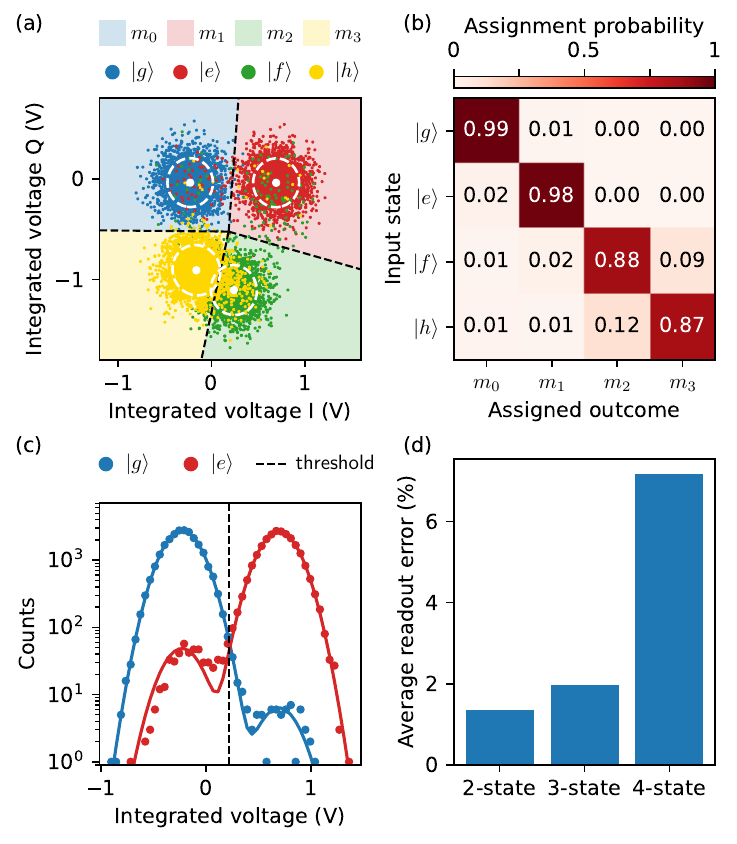}
\caption{\textbf{Four state readout.}
(a) Single-shot readout data for the four lowest-energy transmon states $\ket{g}$, $\ket{e}$, $\ket{f}$ and $\ket{h}$ of $\Ancilla$.
Data are plotted for the first $3\times10^3$ from a total of $2^{15}$ shots for each input state.
The dashed lines show decision boundaries obtained from fitting a linear discrimination classifier to the data.
The mean (white dot) and $3\sigma$ standard deviation (white dashed circles) shown are obtained from Gaussian fits to each input state distribution. (b) Assignment probability matrix obtained from classification of each state into a quaternary outcome. (c) Histogram of shots for qubit states taken along the projection maximizing the signal-to-noise ratio. (d) The average assignment errors for 2-, 3- and 4-state readout are $1.(3)$, $1.(9)$ and $7.(2)\%$, respectively.}
\label{fig:fourstatereadout}
\end{figure}

In order to investigate leakage to higher states in the ancilla, we need to discriminate between the first two leakage states, $\ket{f}$ and $\ket{h}$.
To do this without compromising the performance of the parity check, we simultaneously require high readout fidelity for the qubit states $\ket{g}$ and $\ket{e}$.
We achieve this for the ancilla for the states $\ket{g}$ through $\ket{h}$ using a single readout pulse.
Figure~\ref{fig:fourstatereadout}(a) shows the integrated readout signal for each of the states along with the decision boundaries used to classify the states.
Any leakage to even higher states will likely be assigned to $\ket{h}$ since the resonator response at the readout frequency is mostly flat for $\ket{h}$.
The average assignment error for the four states is $7.(2)\%$ [Fig.~\ref{fig:fourstatereadout}(b)] while the average qubit readout error is $1.(3)\%$ [Fig.~\ref{fig:fourstatereadout}(c)].
Here, we assume that state preparation errors are small compared to assignment errors.

\section{Leakage reduction for higher states}

\begin{figure}
\centering
\includegraphics[width=0.49\textwidth]{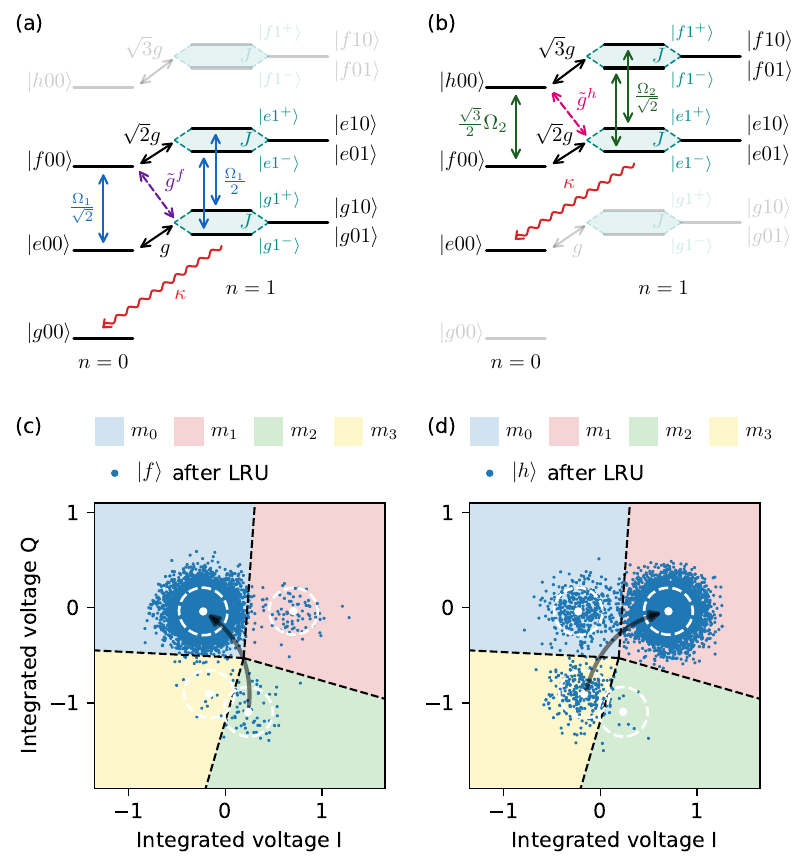}
\caption{\textbf{Leakage reduction for $\ket{f}$ and $\ket{h}$.}
(a, b) Transmon-resonator system level structure showing the relevant couplings for the $f$-LRU (a) and $h$-LRU (b). Each effective coupling, $\tilde{g}^f$ and $\tilde{g}^h$, is mediated by its respective drive $\Omega_1$ and $\Omega_2$ and transmon-resonator coupling $g$. (c, d) Readout data ($2^{13}$ shots) of leakage states $\ket{f}$ (c) and $\ket{h}$ (d) after applying both LRU pulses simultaneously. The white dots and dashed cicles show the mean and $3\sigma$ standard deviation obtained from fitting calibration data for each state.}
\label{fig:h-LRU}
\end{figure}

\begin{figure}
\centering
\includegraphics[width=0.49\textwidth]{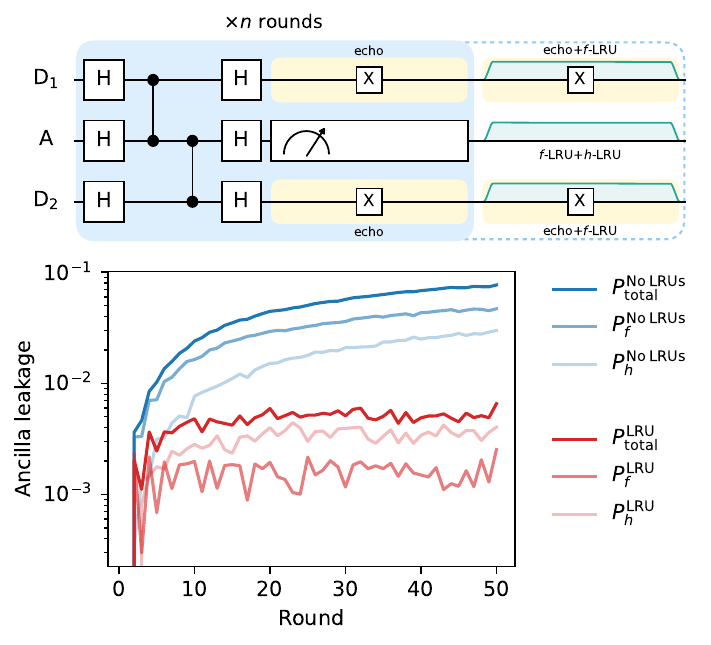}
\caption{\textbf{Higher leakage states of the ancilla qubit.}
Composition of ancilla leakage during the repeated stabilizer measurement of Figure~\ref{fig:repeated_parity_check}. $P_\mathrm{total}=P_f+P_h$ denotes the total leakage population without (blue) and with (red) LRUs.
}
\label{fig:anc_leakage}
\end{figure}

Although the most common leakage mechanisms usually populate the second-excited state of the transmon, $\ket{f}$, some operations such as measurement can leak into higher-excited states~\cite{Sank16}.
We observe the build-up of population in these higher states in the repeated parity-check experiment (Fig.~\ref{fig:repeated_parity_check}).
Figure~\ref{fig:anc_leakage} shows the fraction of total leakage to these higher states for the ancilla.
Therefore, leakage reduction for higher states is necessary for ancillas.
Similar to the leakage reduction mechanism that drives $\ket{f}\rightarrow\ket{g}$ [with effective coupling $\tilde{g}^f$ in Fig.~\ref{fig:h-LRU}(a)], one can drive $\ket{h}\rightarrow\ket{e}$ (with effective coupling $\tilde{g}^h$ in Fig.~\ref{fig:h-LRU}(b)].
This transition can be induced much like the former, with an extra drive at frequency
\begin{equation}
\omega_{h00}-\omega_{e1^\pm} \approx 2\omega_{\mathrm{Q}}+3\alpha-\omega_{\mathrm{RP}},
\label{eq:h-tansition-freq}
\end{equation}
$2\alpha$ below the $f$-LRU transition.
We then have two LRU mechanisms, $f$-LRU and $h$-LRU,  increasing seepage from $\ket{f}$ and $\ket{h}$, respectively.
We drive both of these transitions simultaneously using two independent drives.
Following the same calibration procedure shown in Fig.~\ref{fig:gate_cal} for the $f$-LRU, we tune up a pulse for the $h$-LRU.
Figures~\ref{fig:h-LRU}(c) and~\ref{fig:h-LRU}(d) show readout data for states $\ket{f}$ and $\ket{h}$ after performing both LRUs simultaneously.
The corresponding removal fraction for each state is $R^f=99.(2)\%$ and $R^h=96.(1)\%$ for $\taup=220~\ns$.
Using this scheme, we can effectively reduce leakage in both states (Fig.~\ref{fig:anc_leakage}).
In particular, leakage in $\ket{f}$ is effectively kept under $0.2\%$, while that in $\ket{h}$ sits below $0.4\%$ (red curves in Fig.~\ref{fig:anc_leakage}).
The former shows a flat curve and therefore corresponds to the $L_1$ of the cycle (similar to Fig.~\ref{fig:repeated_LRU_experiment}).
The apparent remaining leakage in $\ket{h}$ could possibly be due to higher-excited states, which are naively assigned as $\ket{h}$ by the readout as they cannot be distinguished.

% Bibliography
% \putbib[../../Paper_resources/References/References_cQED]
%%%%%%%%%%%%%%

%%%%%%%%%%%%%%
\end{bibunit}


\begin{thebibliography}{10}
	\expandafter\ifx\csname url\endcsname\relax
	  \def\url#1{\texttt{#1}}\fi
	\expandafter\ifx\csname urlprefix\endcsname\relax\def\urlprefix{URL }\fi
	\providecommand{\bibinfo}[2]{#2}
	\providecommand{\eprint}[2][]{\url{#2}}

	\bibitem{koch07}
	\bibinfo{author}{Koch, J.} \emph{et~al.}
	\newblock \bibinfo{title}{Charge-insensitive qubit design derived from the
	  {C}ooper pair box}.
	\newblock \emph{\bibinfo{journal}{Phys. Rev. A}} \textbf{\bibinfo{volume}{76}},
	  \bibinfo{pages}{042319} (\bibinfo{year}{2007}).

	\bibitem{Motzoi09}
	\bibinfo{author}{Motzoi, F.}, \bibinfo{author}{Gambetta, J.~M.},
	  \bibinfo{author}{Rebentrost, P.} \& \bibinfo{author}{Wilhelm, F.~K.}
	\newblock \bibinfo{title}{Simple pulses for elimination of leakage in weakly
	  nonlinear qubits}.
	\newblock \emph{\bibinfo{journal}{Phys. Rev. Lett.}}
	  \textbf{\bibinfo{volume}{103}}, \bibinfo{pages}{110501}
	  (\bibinfo{year}{2009}).

	\bibitem{DiCarlo09}
	\bibinfo{author}{Di{C}arlo, L.} \emph{et~al.}
	\newblock \bibinfo{title}{Demonstration of two-qubit algorithms with a
	  superconducting quantum processor}.
	\newblock \emph{\bibinfo{journal}{Nature}} \textbf{\bibinfo{volume}{460}},
	  \bibinfo{pages}{240} (\bibinfo{year}{2009}).

	\bibitem{Barends19}
	\bibinfo{author}{Barends, R.} \emph{et~al.}
	\newblock \bibinfo{title}{Diabatic gates for frequency-tunable superconducting
	  qubits}.
	\newblock \emph{\bibinfo{journal}{Phys. Rev. Lett.}}
	  \textbf{\bibinfo{volume}{123}}, \bibinfo{pages}{210501}
	  (\bibinfo{year}{2019}).

	\bibitem{Negirneac21}
	\bibinfo{author}{Neg\^{\i}rneac, V.} \emph{et~al.}
	\newblock \bibinfo{title}{High-fidelity controlled-$z$ gate with maximal
	  intermediate leakage operating at the speed limit in a superconducting
	  quantum processor}.
	\newblock \emph{\bibinfo{journal}{Phys. Rev. Lett.}}
	  \textbf{\bibinfo{volume}{126}}, \bibinfo{pages}{220502}
	  (\bibinfo{year}{2021}).

	\bibitem{Sank16}
	\bibinfo{author}{Sank, D.} \emph{et~al.}
	\newblock \bibinfo{title}{Measurement-induced state transitions in a
	  superconducting qubit: Beyond the rotating wave approximation}.
	\newblock \emph{\bibinfo{journal}{Phys. Rev. Lett.}}
	  \textbf{\bibinfo{volume}{117}}, \bibinfo{pages}{190503}
	  (\bibinfo{year}{2016}).

	\bibitem{Khezri22}
	\bibinfo{author}{Khezri, M.} \emph{et~al.}
	\newblock \bibinfo{title}{Measurement-induced state transitions in a
	  superconducting qubit: Within the rotating wave approximation}.
	\newblock \emph{\bibinfo{journal}{arXiv preprint arXiv:2212.05097}}
	  (\bibinfo{year}{2022}).

	\bibitem{Chen21}
	\bibinfo{author}{Chen, Z.} \emph{et~al.}
	\newblock \bibinfo{title}{Exponential suppression of bit or phase errors with
	  cyclic error correction}.
	\newblock \emph{\bibinfo{journal}{Nature}} \textbf{\bibinfo{volume}{595}},
	  \bibinfo{pages}{383--387} (\bibinfo{year}{2021}).

	\bibitem{Fowler12}
	\bibinfo{author}{Fowler, A.~G.}, \bibinfo{author}{Mariantoni, M.},
	  \bibinfo{author}{Martinis, J.~M.} \& \bibinfo{author}{Cleland, A.~N.}
	\newblock \bibinfo{title}{Surface codes: Towards practical large-scale quantum
	  computation}.
	\newblock \emph{\bibinfo{journal}{Phys. Rev. A}} \textbf{\bibinfo{volume}{86}},
	  \bibinfo{pages}{032324} (\bibinfo{year}{2012}).

	\bibitem{Riste15}
	\bibinfo{author}{Rist\`{e}, D.} \emph{et~al.}
	\newblock \bibinfo{title}{{Detecting bit-flip errors in a logical qubit using
	  stabilizer measurements}}.
	\newblock \emph{\bibinfo{journal}{Nat.\ Commun.}}
	  \textbf{\bibinfo{volume}{{6}}}, \bibinfo{pages}{6983}
	  (\bibinfo{year}{{2015}}).

	\bibitem{Takita17}
	\bibinfo{author}{Takita, M.}, \bibinfo{author}{Cross, A.~W.},
	  \bibinfo{author}{C\'orcoles, A.~D.}, \bibinfo{author}{Chow, J.~M.} \&
	  \bibinfo{author}{Gambetta, J.~M.}
	\newblock \bibinfo{title}{Experimental demonstration of fault-tolerant state
	  preparation with superconducting qubits}.
	\newblock \emph{\bibinfo{journal}{Phys. Rev. Lett.}}
	  \textbf{\bibinfo{volume}{119}}, \bibinfo{pages}{180501}
	  (\bibinfo{year}{2017}).

	\bibitem{Aliferis07}
	\bibinfo{author}{Aliferis, P.} \& \bibinfo{author}{Terhal, B.~M.}
	\newblock \bibinfo{title}{Fault-tolerant quantum computation for local leakage
	  faults}.
	\newblock \emph{\bibinfo{journal}{Quantum Info. Comput.}}
	  \textbf{\bibinfo{volume}{7}}, \bibinfo{pages}{139--156}
	  (\bibinfo{year}{2007}).

	\bibitem{Fowler13}
	\bibinfo{author}{Fowler, A.~G.}
	\newblock \bibinfo{title}{Coping with qubit leakage in topological codes}.
	\newblock \emph{\bibinfo{journal}{Phys. Rev. A}} \textbf{\bibinfo{volume}{88}},
	  \bibinfo{pages}{042308} (\bibinfo{year}{2013}).

	\bibitem{Ghosh13_b}
	\bibinfo{author}{Ghosh, J.}, \bibinfo{author}{Fowler, A.~G.},
	  \bibinfo{author}{Martinis, J.~M.} \& \bibinfo{author}{Geller, M.~R.}
	\newblock \bibinfo{title}{Understanding the effects of leakage in
	  superconducting quantum-error-detection circuits}.
	\newblock \emph{\bibinfo{journal}{Phys. Rev. A}} \textbf{\bibinfo{volume}{88}},
	  \bibinfo{pages}{062329} (\bibinfo{year}{2013}).

	\bibitem{Varbanov20}
	\bibinfo{author}{Varbanov, B.~M.} \emph{et~al.}
	\newblock \bibinfo{title}{{Leakage detection for a transmon-based surface
	  code}}.
	\newblock \emph{\bibinfo{journal}{npj Quantum Information}}
	  \textbf{\bibinfo{volume}{6}}, \bibinfo{pages}{102} (\bibinfo{year}{2020}).

	\bibitem{Bultink20}
	\bibinfo{author}{Bultink, C.~C.} \emph{et~al.}
	\newblock \bibinfo{title}{Protecting quantum entanglement from leakage and
	  qubit errors via repetitive parity measurements}.
	\newblock \emph{\bibinfo{journal}{Science Advances}}
	  \textbf{\bibinfo{volume}{6}} (\bibinfo{year}{2020}).

	\bibitem{Mcewen21}
	\bibinfo{author}{McEwen, M.} \emph{et~al.}
	\newblock \bibinfo{title}{Removing leakage-induced correlated errors in
	  superconducting quantum error correction}.
	\newblock \emph{\bibinfo{journal}{Nature communications}}
	  \textbf{\bibinfo{volume}{12}}, \bibinfo{pages}{1--7} (\bibinfo{year}{2021}).

	\bibitem{Ryan-Anderson21}
	\bibinfo{author}{Ryan-Anderson, C.} \emph{et~al.}
	\newblock \bibinfo{title}{Realization of real-time fault-tolerant quantum error
	  correction}.
	\newblock \emph{\bibinfo{journal}{Phys. Rev. X}} \textbf{\bibinfo{volume}{11}},
	  \bibinfo{pages}{041058} (\bibinfo{year}{2021}).

	\bibitem{krinner22}
	\bibinfo{author}{Krinner, S.} \emph{et~al.}
	\newblock \bibinfo{title}{Realizing repeated quantum error correction in a
	  distance-three surface code}.
	\newblock \emph{\bibinfo{journal}{Nature}} \textbf{\bibinfo{volume}{605}},
	  \bibinfo{pages}{669--674} (\bibinfo{year}{2022}).

	\bibitem{Zhao22}
	\bibinfo{author}{Zhao, Y.} \emph{et~al.}
	\newblock \bibinfo{title}{Realization of an error-correcting surface code with
	  superconducting qubits}.
	\newblock \emph{\bibinfo{journal}{Phys. Rev. Lett.}}
	  \textbf{\bibinfo{volume}{129}}, \bibinfo{pages}{030501}
	  (\bibinfo{year}{2022}).

	\bibitem{Sundaresan22}
	\bibinfo{author}{Sundaresan, N.} \emph{et~al.}
	\newblock \bibinfo{title}{Matching and maximum likelihood decoding of a
	  multi-round subsystem quantum error correction experiment}.
	\newblock \emph{\bibinfo{journal}{arXiv preprint arXiv:2203.07205}}
	  (\bibinfo{year}{2022}).

	\bibitem{Acharya22}
	\bibinfo{author}{Acharya, R.} \emph{et~al.}
	\newblock \bibinfo{title}{Suppressing quantum errors by scaling a surface code
	  logical qubit}.
	\newblock \emph{\bibinfo{journal}{arXiv preprint arXiv:2207.06431}}
	  (\bibinfo{year}{2022}).

	\bibitem{Postler22}
	\bibinfo{author}{Postler, L.} \emph{et~al.}
	\newblock \bibinfo{title}{Demonstration of fault-tolerant universal quantum
	  gate operations}.
	\newblock \emph{\bibinfo{journal}{Nature}} \textbf{\bibinfo{volume}{605}},
	  \bibinfo{pages}{675--680} (\bibinfo{year}{2022}).

	\bibitem{Magnard18}
	\bibinfo{author}{Magnard, P.} \emph{et~al.}
	\newblock \bibinfo{title}{Fast and unconditional all-microwave reset of a
	  superconducting qubit}.
	\newblock \emph{\bibinfo{journal}{Phys. Rev. Lett.}}
	  \textbf{\bibinfo{volume}{121}}, \bibinfo{pages}{060502}
	  (\bibinfo{year}{2018}).

	\bibitem{Riste12}
	\bibinfo{author}{Rist\`e, D.}, \bibinfo{author}{van Leeuwen, J.~G.},
	  \bibinfo{author}{Ku, H.-S.}, \bibinfo{author}{Lehnert, K.~W.} \&
	  \bibinfo{author}{DiCarlo, L.}
	\newblock \bibinfo{title}{Initialization by measurement of a superconducting
	  quantum bit circuit}.
	\newblock \emph{\bibinfo{journal}{Phys. Rev. Lett.}}
	  \textbf{\bibinfo{volume}{109}}, \bibinfo{pages}{050507}
	  (\bibinfo{year}{2012}).

	\bibitem{Andersen19}
	\bibinfo{author}{Andersen, C.~K.} \emph{et~al.}
	\newblock \bibinfo{title}{Entanglement stabilization using ancilla-based parity
	  detection and real-time feedback in superconducting circuits}.
	\newblock \emph{\bibinfo{journal}{npj Quantum Information}}
	  \textbf{\bibinfo{volume}{5}}, \bibinfo{pages}{1--7} (\bibinfo{year}{2019}).

	\bibitem{Ghosh15}
	\bibinfo{author}{Ghosh, J.} \& \bibinfo{author}{Fowler, A.~G.}
	\newblock \bibinfo{title}{Leakage-resilient approach to fault-tolerant quantum
	  computing with superconducting elements}.
	\newblock \emph{\bibinfo{journal}{Phys. Rev. A}} \textbf{\bibinfo{volume}{91}},
	  \bibinfo{pages}{020302(R)} (\bibinfo{year}{2015}).

	\bibitem{Suchara15}
	\bibinfo{author}{Suchara, M.}, \bibinfo{author}{Cross, A.~W.} \&
	  \bibinfo{author}{Gambetta, J.~M.}
	\newblock \bibinfo{title}{Leakage suppression in the toric code}.
	\newblock \emph{\bibinfo{journal}{Quantum Info. Comput.}}
	  \textbf{\bibinfo{volume}{15}}, \bibinfo{pages}{997--1016}
	  (\bibinfo{year}{2015}).

	\bibitem{Mcewen23}
	\bibinfo{author}{McEwen, M.} \emph{et~al.}
	\newblock \bibinfo{title}{Relaxing Hardware Requirements for Surface Code Circuits
	  using Time-dynamics}.
	\newblock \emph{\bibinfo{journal}{arXiv preprint arXiv:2302.02192}}
	  (\bibinfo{year}{2023}).

	\bibitem{Brown19}
	\bibinfo{author}{Brown, N.~C.}, \bibinfo{author}{Newman, M.} \&
	  \bibinfo{author}{Brown, K.~R.}
	\newblock \bibinfo{title}{Handling leakage with subsystem codes}.
	\newblock \emph{\bibinfo{journal}{New Journal of Physics}}
	  \textbf{\bibinfo{volume}{21}}, \bibinfo{pages}{073055}
	  (\bibinfo{year}{2019}).

	\bibitem{Brown20}
	\bibinfo{author}{Brown, N.~C.}, \bibinfo{author}{Cross, A.} \&
	  \bibinfo{author}{Brown, K.~R.}
	\newblock \bibinfo{title}{Critical faults of leakage errors on the surface
	  code}.
	\newblock In \emph{\bibinfo{booktitle}{2020 IEEE International Conference on
	  Quantum Computing and Engineering (QCE)}}, \bibinfo{pages}{286--294}
	  (\bibinfo{year}{2020}).

	\bibitem{Hayes20}
	\bibinfo{author}{Hayes, D.} \emph{et~al.}
	\newblock \bibinfo{title}{Eliminating leakage errors in hyperfine qubits}.
	\newblock \emph{\bibinfo{journal}{Phys. Rev. Lett.}}
	  \textbf{\bibinfo{volume}{124}}, \bibinfo{pages}{170501}
	  (\bibinfo{year}{2020}).

	\bibitem{Battistel21}
	\bibinfo{author}{Battistel, F.}, \bibinfo{author}{Varbanov, B.} \&
	  \bibinfo{author}{Terhal, B.}
	\newblock \bibinfo{title}{Hardware-efficient leakage-reduction scheme for
	  quantum error correction with superconducting transmon qubits}.
	\newblock \emph{\bibinfo{journal}{PRX Quantum}} \textbf{\bibinfo{volume}{2}},
	  \bibinfo{pages}{030314} (\bibinfo{year}{2021}).

	\bibitem{Miao22}
	\bibinfo{author}{Miao, K.~C.} \emph{et~al.}
	\newblock \bibinfo{title}{Overcoming leakage in scalable quantum error
	  correction}.
	\newblock \emph{\bibinfo{journal}{arXiv preprint arXiv:2211.04728}}
	  (\bibinfo{year}{2022}).

	\bibitem{Grassl97}
	\bibinfo{author}{Grassl, M.}, \bibinfo{author}{Beth, T.} \&
	  \bibinfo{author}{Pellizzari, T.}
	\newblock \bibinfo{title}{Codes for the quantum erasure channel}.
	\newblock \emph{\bibinfo{journal}{Phys. Rev. A}} \textbf{\bibinfo{volume}{56}},
	  \bibinfo{pages}{33--38} (\bibinfo{year}{1997}).

	\bibitem{Bennett97}
	\bibinfo{author}{Bennett, C.~H.}, \bibinfo{author}{DiVincenzo, D.~P.} \&
	  \bibinfo{author}{Smolin, J.~A.}
	\newblock \bibinfo{title}{Capacities of quantum erasure channels}.
	\newblock \emph{\bibinfo{journal}{Phys. Rev. Lett.}}
	  \textbf{\bibinfo{volume}{78}}, \bibinfo{pages}{3217--3220}
	  (\bibinfo{year}{1997}).

	\bibitem{Stace09}
	\bibinfo{author}{Stace, T.~M.}, \bibinfo{author}{Barrett, S.~D.} \&
	  \bibinfo{author}{Doherty, A.~C.}
	\newblock \bibinfo{title}{Thresholds for topological codes in the presence of
	  loss}.
	\newblock \emph{\bibinfo{journal}{Phys. Rev. Lett.}}
	  \textbf{\bibinfo{volume}{102}}, \bibinfo{pages}{200501}
	  (\bibinfo{year}{2009}).

	\bibitem{Barrett10}
	\bibinfo{author}{Barrett, S.~D.} \& \bibinfo{author}{Stace, T.~M.}
	\newblock \bibinfo{title}{Fault tolerant quantum computation with very high
	  threshold for loss errors}.
	\newblock \emph{\bibinfo{journal}{Phys. Rev. Lett.}}
	  \textbf{\bibinfo{volume}{105}}, \bibinfo{pages}{200502}
	  (\bibinfo{year}{2010}).

	\bibitem{Kubica22}
	\bibinfo{author}{Kubica, A.}, \bibinfo{author}{Haim, A.},
	  \bibinfo{author}{Vaknin, Y.}, \bibinfo{author}{Brandão, F.} \&
	  \bibinfo{author}{Retzker, A.}
	\newblock \bibinfo{title}{Erasure qubits: Overcoming the $T_1$ limit in
	  superconducting circuits} (\bibinfo{year}{2022}).

	\bibitem{Wu2022}
	\bibinfo{author}{Wu, Y.}, \bibinfo{author}{Kolkowitz, S.},
	  \bibinfo{author}{Puri, S.} \& \bibinfo{author}{Thompson, J.~D.}
	\newblock \bibinfo{title}{Erasure conversion for fault-tolerant quantum
	  computing in alkaline earth Rydberg atom arrays}.
	\newblock \emph{\bibinfo{journal}{Nature Communications}}
	  \textbf{\bibinfo{volume}{13}}, \bibinfo{pages}{4657} (\bibinfo{year}{2022}).

	\bibitem{Heinsoo18}
	\bibinfo{author}{Heinsoo, J.} \emph{et~al.}
	\newblock \bibinfo{title}{Rapid high-fidelity multiplexed readout of
	  superconducting qubits}.
	\newblock \emph{\bibinfo{journal}{Phys. Rev. App.}}
	  \textbf{\bibinfo{volume}{10}}, \bibinfo{pages}{034040}
	  (\bibinfo{year}{2018}).

	\bibitem{Chow12}
	\bibinfo{author}{Chow, J.~M.} \emph{et~al.}
	\newblock \bibinfo{title}{Universal quantum gate set approaching fault-tolerant
	  thresholds with superconducting qubits}.
	\newblock \emph{\bibinfo{journal}{Phys. Rev. Lett.}}
	  \textbf{\bibinfo{volume}{109}}, \bibinfo{pages}{060501}
	  (\bibinfo{year}{2012}).

	\bibitem{Versluis17}
	\bibinfo{author}{Versluis, R.} \emph{et~al.}
	\newblock \bibinfo{title}{Scalable quantum circuit and control for a
	  superconducting surface code}.
	\newblock \emph{\bibinfo{journal}{Phys. Rev. App.}}
	  \textbf{\bibinfo{volume}{8}}, \bibinfo{pages}{034021} (\bibinfo{year}{2017}).
\end{thebibliography}

\begin{thebibliography}{10}
	\expandafter\ifx\csname url\endcsname\relax
	  \def\url#1{\texttt{#1}}\fi
	\expandafter\ifx\csname urlprefix\endcsname\relax\def\urlprefix{URL }\fi
	\providecommand{\bibinfo}[2]{#2}
	\providecommand{\eprint}[2][]{\url{#2}}

	\bibitem{Krantz19}
	\bibinfo{author}{Krantz, P.} \emph{et~al.}
	\newblock \bibinfo{title}{A quantum engineer's guide to superconducting
	  qubits}.
	\newblock \emph{\bibinfo{journal}{App. Phys. Rev.}}
	  \textbf{\bibinfo{volume}{6}}, \bibinfo{pages}{021318} (\bibinfo{year}{2019}).

	\bibitem{Magesan12}
	\bibinfo{author}{Magesan, E.}, \bibinfo{author}{Gambetta, J.~M.} \&
	  \bibinfo{author}{Emerson, J.}
	\newblock \bibinfo{title}{Characterizing quantum gates via randomized
	  benchmarking}.
	\newblock \emph{\bibinfo{journal}{Phys. Rev. A}} \textbf{\bibinfo{volume}{85}},
	  \bibinfo{pages}{042311} (\bibinfo{year}{2012}).

	\bibitem{Magesan12b}
	\bibinfo{author}{Magesan, E.} \emph{et~al.}
	\newblock \bibinfo{title}{Efficient measurement of quantum gate error by
	  interleaved randomized benchmarking}.
	\newblock \emph{\bibinfo{journal}{Phys. Rev. Lett.}}
	  \textbf{\bibinfo{volume}{109}}, \bibinfo{pages}{080505}
	  (\bibinfo{year}{2012}).

	\bibitem{Wood18}
	\bibinfo{author}{Wood, C.~J.} \& \bibinfo{author}{Gambetta, J.~M.}
	\newblock \bibinfo{title}{Quantification and characterization of leakage
	  errors}.
	\newblock \emph{\bibinfo{journal}{Phys. Rev. A}} \textbf{\bibinfo{volume}{97}},
	  \bibinfo{pages}{032306} (\bibinfo{year}{2018}).

	\bibitem{Versluis17}
	\bibinfo{author}{Versluis, R.} \emph{et~al.}
	\newblock \bibinfo{title}{Scalable quantum circuit and control for a
	  superconducting surface code}.
	\newblock \emph{\bibinfo{journal}{Phys. Rev. App.}}
	  \textbf{\bibinfo{volume}{8}}, \bibinfo{pages}{034021} (\bibinfo{year}{2017}).

	\bibitem{Heinsoo18}
	\bibinfo{author}{Heinsoo, J.} \emph{et~al.}
	\newblock \bibinfo{title}{Rapid high-fidelity multiplexed readout of
	  superconducting qubits}.
	\newblock \emph{\bibinfo{journal}{Phys. Rev. App.}}
	  \textbf{\bibinfo{volume}{10}}, \bibinfo{pages}{034040}
	  (\bibinfo{year}{2018}).

	\bibitem{Motzoi13}
	\bibinfo{author}{Motzoi, F.} \& \bibinfo{author}{Wilhelm, F.~K.}
	\newblock \bibinfo{title}{Improving frequency selection of driven pulses using
	  derivative-based transition suppression}.
	\newblock \emph{\bibinfo{journal}{Phys. Rev. A}} \textbf{\bibinfo{volume}{88}},
	  \bibinfo{pages}{062318} (\bibinfo{year}{2013}).

	\bibitem{Negirneac21}
	\bibinfo{author}{Neg\^{\i}rneac, V.} \emph{et~al.}
	\newblock \bibinfo{title}{High-fidelity controlled-$z$ gate with maximal
	  intermediate leakage operating at the speed limit in a superconducting
	  quantum processor}.
	\newblock \emph{\bibinfo{journal}{Phys. Rev. Lett.}}
	  \textbf{\bibinfo{volume}{126}}, \bibinfo{pages}{220502}
	  (\bibinfo{year}{2021}).

	\bibitem{Sank16}
	\bibinfo{author}{Sank, D.} \emph{et~al.}
	\newblock \bibinfo{title}{Measurement-induced state transitions in a
	  superconducting qubit: Beyond the rotating wave approximation}.
	\newblock \emph{\bibinfo{journal}{Phys. Rev. Lett.}}
	  \textbf{\bibinfo{volume}{117}}, \bibinfo{pages}{190503}
	  (\bibinfo{year}{2016}).

	\bibitem{Chow12}
	\bibinfo{author}{Chow, J.~M.} \emph{et~al.}
	\newblock \bibinfo{title}{Universal quantum gate set approaching fault-tolerant
	  thresholds with superconducting qubits}.
	\newblock \emph{\bibinfo{journal}{Phys. Rev. Lett.}}
	  \textbf{\bibinfo{volume}{109}}, \bibinfo{pages}{060501}
	  (\bibinfo{year}{2012}).

	\bibitem{Khezri22}
	\bibinfo{author}{Khezri, M.} \emph{et~al.}
	\newblock \bibinfo{title}{Measurement-induced state transitions in a
	  superconducting qubit: Within the rotating wave approximation}.
	\newblock \emph{\bibinfo{journal}{arXiv preprint arXiv:2212.05097}}
	  (\bibinfo{year}{2022}).
\end{thebibliography}
\end{document}